# Asymmetric topological pumping in nonparaxial photonics

Qingqing Cheng[1], Huaiqiang Wang[2,3], Yongguan Ke[4,5], Tao Chen[1], Ye Yu[1], Yuri S. Kivshar[6✉], Chaohong Lee[4,5✉] & Yiming Pan[7✉]

Topological photonics was initially inspired by the quantum-optical analogy between the Schrödinger equation for an electron wavefunction and the paraxial equation for a light beam. Here, we reveal an unexpected phenomenon in topological pumping observed in arrays of nonparaxial optical waveguides where the quantum-optical analogy becomes invalid. We predict theoretically and demonstrate experimentally an asymmetric topological pumping when the injected field transfers from one side of the waveguide array to the other side whereas the reverse process is unexpectedly forbidden. Our finding could open an avenue for exploring topological photonics that enables nontrivial topological phenomena and designs in photonics driven by nonparaxiality.

[1] Shanghai Key Laboratory of Modern Optical System, School of Optical-Electrical and Computer Engineering, University of Shanghai for Science and Technology, Shanghai 200093, China. [2] National Laboratory of Solid State Microstructures, School of Physics, Nanjing University, Nanjing 210093, China. [3] Department of Physics, University of Zurich, 8057 Zurich, Switzerland. [4] Guangdong Provincial Key Laboratory of Quantum Metrology and Sensing & School of Physics and Astronomy, Sun Yat-Sen University (Zhuhai Campus), Zhuhai 519082, China. [5] State Key Laboratory of Optoelectronic Materials and Technologies, Sun Yat-Sen University (Guangzhou Campus), Guangzhou 510275, China. [6] Nonlinear Physics Centre, Research School of Physics, Australian National University, Canberra, ACT 2601, Australia. [7] Physics Department and Solid State Institute, Technion, Haifa 32000, Israel. ✉email: yuri.kivshar@anu.edu.au; lichaoh2@mail.sysu.edu.cn; yiming.pan@campus.technion.ac.il

The field of topological photonics[1–5] emerged after the discovery of topological insulators for electrons[6,7], and it utilizes topology and topological phases to tailor behaviors of light in an unusual way. Seminal studies in topological photonics include quantum simulations of topological edge states[8–14] and topological pumping[15–21]. Similarities of those concepts are based primarily on a direct correspondence between the dynamics of the electron wavefunction governed by the Schrödinger equation and the propagation of a light beam described by the paraxial Helmholtz equation. Beyond quantum simulations, unusual topological effects in photonics are generated by intrinsic non-Hermiticity such as topological lasers[22–27], nonlinear response including active tunability[25,26,28], nonreciprocity[29,30], and frequency conversion[31,32].

Of particular interest is the topological pumping of light beams in periodically-modulated waveguide arrays[9,16,33–35], which may find applications in quantum information processing, such as preparing entangled states of photons[36,37] and transferring quantum states[38–41]. The first topological pumping scheme is known as Thouless pumping[42–47], and it was proposed and realized in photonic systems[48–51], in which the light field fills uniformly a bulk band and transfers by integer unit cells. An alternative topological pumping is to transfer a localized light field from one side to another, realized in optical frequency regime[16]. Up to now, the topological pumping of light was studied in the framework of quantum simulations, where specific optical properties do not play any role[38,52–55]. In the low-frequency regime, the nonparaxial light propagation cannot be simply mapped to a Schrödinger-type wave equation. For example, a recent study[56] revealed that a tightly focused nonparaxial light beam can enhance nonlinear optical interactions. However, no attempt has been done to explore the topological properties of nonparaxial light.

Here, we consider the case when the quantum-optical analogy is no longer valid, and when the paraxiality requirement is removed. We demonstrate experimentally a remarkable effect of asymmetric topological pumping when the injected field transfers from one side of the waveguide array to the other side, whereas the reverse process becomes forbidden. Intriguingly, the pump setup of the light field is designed and fabricated by varying the waveguide structural configurations according to the celebrated Rice-Mele model[57], which predicts the symmetric topological pumping. The key to understanding this effect of asymmetric topological pumping is that an effective negative next-nearest-neighboring (NNN) coupling is induced by nonparaxiality, which levitates the symmetric spectral spacing of the energy spectrum of the paraxial light beam. This spectral levitation allows to maintain one pump channel but block the other channel by obeying and violating adiabatic conditions, respectively, in an asymmetric fashion. Notably, the negative NNN coupling cannot arise from the direct coupling between waveguides[58], but instead it stems from the nonparaxial beam propagation of microwaves. Indeed, our experimental results are nicely consistent with both simulations and analysis from the exact Helmholtz wave equation. We believe these findings may open an avenue to topological photonics, and thus may provide further insights for the study of topological phases and dynamics triggered by nonparaxiality of light.

## Results

**Model**. To illustrate the topological photonic pumping in our microwave waveguide platform, we take advantage of the celebrated Rice-Mele model[57,59]. Two essential factors of the Rice-Mele pump setup are time-periodic staggered on-site potentials ($\beta_{1,2}$) and dimerized nearest-neighbor (NN) hoppings ($\kappa_{1,2}$), as shown schematically in Fig. 1a, b. By regulating the periodically curved waveguides into an array[60], we can construct the time-dependent NN hopping along the propagation direction $z$, where the $z$-direction acts as the synthetic time [see Fig. 1a]. The coupling constants between the $j$th and $(j+1)$th waveguides can be approximated as $\kappa_j(z) \approx \kappa_0 + (-1)^j \delta\kappa \cos(2\pi z/\Lambda)$ (see Fig. S1 for details). As to the periodically staggered on-site potential energy, the structural parameters of the waveguide vary along the $z$-direction to make the time-dependence of the propagation constant follow a sinusoidal function. The Rice-Mele setting can thus be performed by the flexible designs[61,62] of the periodic "H-shaped" structural unit imprinted on the microwave waveguides. Here, we alter the height parameter of the "H-shaped" unit as a sinusoidal function of the propagation direction $z$ to obtain the periodically-modulated on-site potential in the Rice-Mele setting (see Fig. S2 for more details), leading to the desired on-site potentials $\beta_j(z) \approx \beta_0 + (-1)^j \delta\beta \sin(2\pi z/\Lambda)$. Via the above grooves-corrugated waveguide design, our coupled waveguide array can offer us a photonic analog of the Rice-Mele model (as shown in Fig. S3), in which the corresponding tight-binding Hamiltonian of guiding mode amplitudes $\hat{c}_j$ can be described as:

$$\begin{aligned}H_{\mathrm{RM}} = &\sum_{j=1}^{N} \left[\beta_0 + (-1)^j \delta\beta \sin(2\pi z/\Lambda)\right] \hat{c}_j^\dagger \hat{c}_j \\ &+ \sum_{j=1}^{N-1} \left[\kappa_0 + (-1)^j \delta\kappa \cos(2\pi z/\Lambda)\right] \hat{c}_j^\dagger \hat{c}_{j+1} + \mathrm{H.c.}.\end{aligned} \quad (1)$$

For our setup, the experimental parameters are $N = 10$, $\Lambda = 42$ cm, $\kappa_0 = 0.034$ mm$^{-1}$, $\delta\kappa = 0.021$ cm$^{-1}$, $\beta_0 = 0.47$ mm$^{-1}$, and $\delta\beta = 0.03$ mm$^{-1}$, and the distributions of the functions $\beta_{1,2}(z)$ and $\kappa_{1,2}(z)$ are explicitly presented in Fig. 1c, d, respectively. More specifically, $\kappa_{1,2}(z)$ depends not only on the spacings between waveguides, but also on the geometry of spoof surface plasmon polariton (SPP) structures. Therefore, the precise coupling function in Fig. 1d is a complicated one, as shown in Fig. S4.

In such periodically-modulated waveguide arrays, the pump dynamics is governed by the function $|\psi(z)\rangle = \mathcal{T} \exp(i \int_0^z H_{\mathrm{RM}}(z') dz') |\psi(0)\rangle$ with the time-ordering operator $\mathcal{T}$. Microwaves injected from one side should be transported to the other side after a pump cycle, that is to say, the initial state $\psi(0)$ is an instantaneous zero mode at one edge, and the final state $\psi(T)$ becomes the other zero mode at the other edge after one cyclic evolution. Three points need to be emphasized for this prediction before proceeding to our experimental results. First, the paraxial approximation is taken into account to derive the above Hamiltonian within the coupled-mode theory. Second, the adiabatic pumping requires the adiabatic condition that the modulation frequency is much smaller than the minimal spectral spacing between the pump channel and bulk states. Third, in the pump process from the left boundary $|\psi_{L \to R}(z)\rangle$, and the reverse process from the right boundary $|\psi_{R \to L}(z)\rangle$, the spatial probability distributions obey the relation $|\psi_{L \to R}(z, j)|^2 = |\psi_{R \to L}(z, N+1-j)|^2$, indicating that the pumping from the left to right edges should be symmetric to that from the right to left.

**Experimental results**. In our photonic Rice-Mele setting consisting of ten spoof SPP waveguides, microwaves are injected from the left (the 1st) and the right (the 10th) boundary waveguides to excite two pump channels, respectively, i.e., one from left to right and the counterpart one from right to left. We then carry out experimental near-field measurements of the field distribution of the electric component $E_y$ (see Fig. S5). We find that when injecting 17 GHz microwaves from the 1st waveguide, the near-field distribution $E_y$ shows that the microwave gradually transfers to the 10th waveguide, as like the microwave spreads into the bulk and converges again on the right boundary of the array [see Fig. 1e], which indeed confirms with the left-to-right topological pump process. However, as shown in Fig. 1f, when this microwave is injected from the 10th waveguide, the propagation dynamics exhibits some scattering into the bulk waveguides and resides mainly near the right boundary waveguide without further transferring to the left boundary, reflecting the breakdown of the right-to-left pump channel. This failed pumping obviously contradicts the symmetric pumping prediction in our well-designed Rice-Mele setup as mentioned above. To this end, the observation of this dramatic asymmetric pump behavior is one of the main achievements of our work.

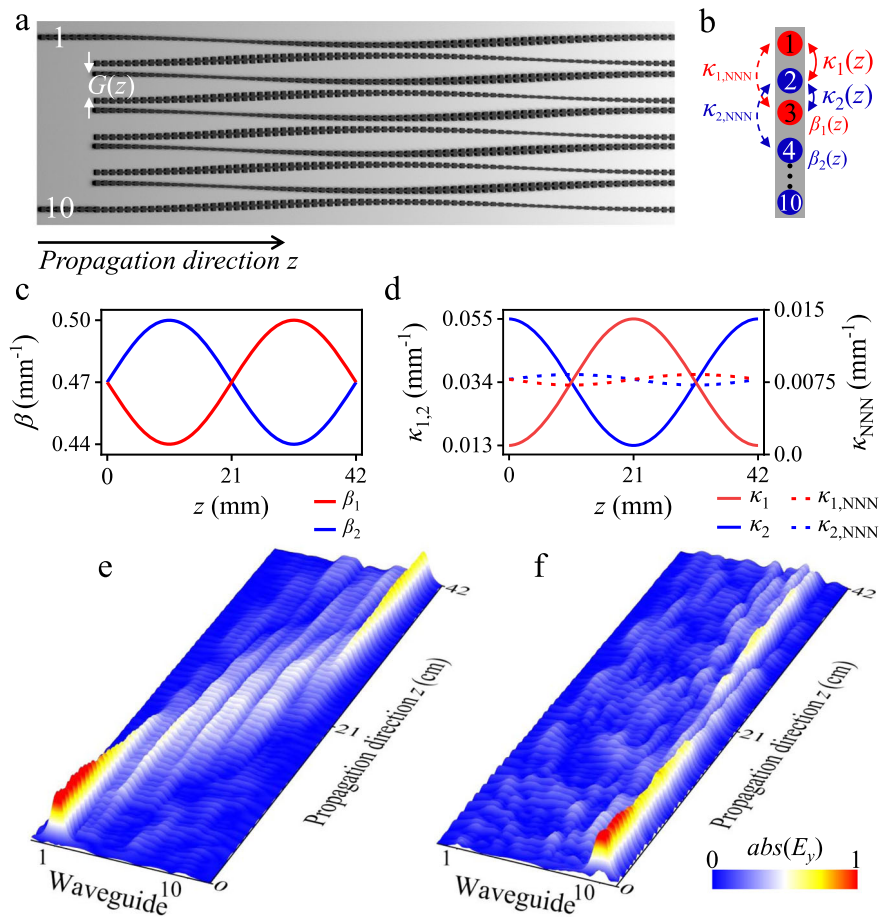

**Fig. 1 Asymmetric topological pumping in an array of SPPs waveguides. a** Schematic illustration of the photonic Rice-Mele model consisting of periodically bending waveguides, with sinusoidal modulations of the nearest-neighbor (NN) waveguide spacing $G_z$ $G(z)$ and height parameters of the "H-shaped" unit along the propagation direction $z$. **b** Illustration of the extended Rice-Mele model with staggered on-site potentials (propagation constants $\beta_{1,2}$), NN coupling strengths ($\kappa_{1,2}$), and next-nearest-neighbor coupling strength ($\kappa_{NNN}$). The detailed values of the sinusoidally modulated (**c**) propagating constants and (**d**) coupling strengths $\kappa_{NN}$ and $\kappa_{NNN}$ along the propagation direction $z$. Experimental data for the near-field measurements of the electric field ($E_y$, perpendicular to the array surface) after injecting 17 GHz microwaves from **e** left boundary (the 1st waveguide) and **f** right boundary (the 10th waveguide), respectively.

**Energy spectrum and beam propagation**. To explain the unexpected experimental observation of asymmetric pumping, an intuitive understanding suggests that the Rice-Mele Hamiltonian (1) is insufficient to fully describe the pump process. The Hamiltonian can be amended by adding the most natural term from the inevitable direct NNN coupling ($\kappa_{NNN}$) between the same sub-lattices [see Fig. 1(b)]. We notice that the direct NNN coupling should be always positive, as similar to the nearest-neighboring coupling profile [see Fig. 1d]. However, as we demonstrate below, such a positive NNN coupling still insufficient to explain the experimental data.

To resolve the contradiction between the observed asymmetric pumping and the modified Rice-Mele model, we start by inspecting the particle-hole symmetric case with $\kappa_{NNN} = 0$, as shown in Fig. 2a by the tight-binding band structure of the 10-waveguide system as a function of the phase parameter $\phi \equiv 2\pi z/\Lambda$. Pump channels from the left to right (red solid line) and from the right to left (blue dashed line) can be seen clearly in the bulk-band gap, which is symmetric mutually with the same magnitude and opposite sign of the energy dispersion. In Fig. 2d, by explicitly calculating the instantaneous wavefunction distribution at each values of $\phi$, we simulate the propagating behaviors of microwaves injected from the left and right boundary waveguides, respectively. Obviously, both pump channels are unblocked with symmetric propagating patterns, and microwaves can thus be pumped from left to right and vice versa.

However, in the presence of a positive direct NNN coupling $\kappa_{NNN} > 0$, which can be approximated by a constant (~0.007 mm$^{-1}$) with small fluctuations depending on $\phi$, the band structure is no longer symmetric, as shown in Fig. 2b. The bandwidth and overall level spacing of the valence (conduction) band at each $\phi$ get significantly compressed (enlarged). Notably, the spectral spacing between the left-to-right pump channel (red solid line) and the valence band is severely compressed, while the spectral spacing between the right-to-left pump channel (blue dashed line) and the conduction band is enlarged. According to the Landau-Zener-Stückelberg transition process[63], the adiabatic condition would presumably be violated for the left-to-right pumping because of the severely-concentrated spectral spacing, but it remains satisfied for the right-to-left pumping. The asymmetric breakdown of the adiabatic condition further leads to the scattering of the light field in the left-to-right pump channel to the bulk, and thus blocks this channel. As illustrated in Fig. 2e, the numerical simulations indicate that the left-to-right pumping fails whereas the right-to-left pumping holds. Although the asymmetric pumping could emerge with the positive direct NNN coupling in the lower panel of Fig. 2e, it runs counter to the experimental observation in Fig. 1f, in which the right-to-left pump channel is blocked.

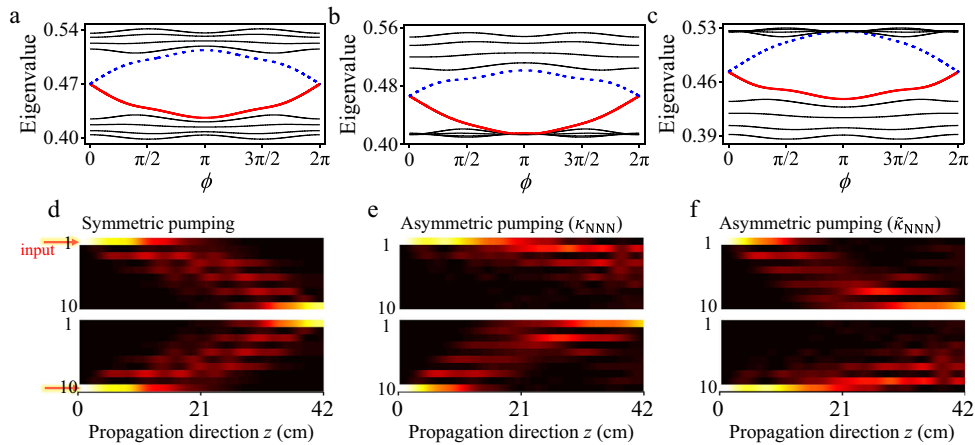

**Fig. 2 Energy spectrum and beam propagation for topological pumping.** The energy spectrum for **a** conventional Rice-Mele model without next-nearest-neighbor (NNN) hopping, and **b**, **c** extended Rice-Mele model with positive ($\kappa_{NNN} = 0.007\,\text{mm}^{-1}$) and negative ($\tilde{\kappa}_{NNN} = -0.007\,\text{mm}^{-1}$) NNN hopping terms, respectively. Two symmetric pump channels are shown in **a** but only one of the channels (**b**, **c**) remains when adding the NNN hopping. Correspondingly, the beam propagation on these photonic lattices demonstrate two symmetric pumping in **d** and asymmetric pumping in **e**, **f**. The beam propagation in the blocked channels [the upper panel in **e**, and the lower panel in **f**] show the topological edge state confinement along the array boundaries. The simulation parameters are taken from the experiment.

Therefore, in light of the above discussion, if the NNN coupling is negative $\tilde{\kappa}_{NNN} < 0$, in contrast to the positive case, we find that the spectral spacing between left-to-right (right-to-left) pump channel and valence (conduction) bands becomes enlarged (compressed), as shown in Fig. 2c. As a result, this leads to the compelling asymmetric pumping [Fig. 2f], in consistent with the experimental observations.

## Discussion
Now, a natural question arises that how the negative NNN coupling appears persuasively in the nonparaxial waveguide system? We recall that the paraxial approximation was adopted in the effective photonic tight-binding Hamiltonian (1) through the coupled-mode theory, which requires the stringent condition $|\partial^2\psi/\partial z^2| \ll 2k|\partial\psi/\partial z|$. Here, $\psi$ is the field distribution of the electric component (e.g., $E_y$), $k$ is the wavevector. However, for the microwave system with much smaller wavevectors, this condition is not satisfied, and we need to go beyond the coupled-mode theory and start from the original exact two dimensional Helmholtz equation of the waveguide systems,

$$i\frac{\partial\psi}{\partial z} - \frac{1}{2kn_0}\frac{\partial^2\psi}{\partial z^2} = H\psi, \quad (2)$$

where $H = \frac{1}{2kn_0}\nabla_x^2 + \frac{k}{2}\left(\frac{n^2-n_0^2}{n_0}\right)$ is the defined Helmholtz-Hamiltonian operator, $n_0$ is reference refractive index, and for monochromatic electromagnetic wave $E(x,z;t) = \psi(x)e^{i\omega t - ikn_0 z}$, the wavevector is $k = \omega/c$. Indeed, solving the exact Helmholtz equation can provide almost the same results as those observed experimentally (see Fig. S6 for details).

To understand qualitatively the origin of the negative NNN coupling, we replace the Helmholtz-Hamiltonian operator via the geometric modeling of the Rice-Mele Hamiltonian $H_{RM}$ in our system, and rewrite Eq. (2) in an effective Schrödinger-type form

$$i\frac{\partial\psi}{\partial z} = \frac{H_{RM}}{1 + i\frac{1}{2kn_0}\frac{\partial}{\partial z}}\psi = H_{eff}\psi. \quad (3)$$

In Eq. (2) the nonparaxial term, contributes to this iteration form, which can be formally expressed as an effective $H_{eff}$ by using many perturbation methods. Full details of Eq. (2) and Eq. (3) are provided in Supplementary Materials. It should be emphasized that the above iterative equation remains exact beyond the paraxial approximation, and $H_{eff}$ can be treated by the known Padé approximation[58]. The Padé approximation can provide an effective Hamiltonian $H_{eff}^{(m,n)} = N(m)/D(n)$, where $N(m)$, and $D(n)$ are polynomials of $H_{RM}$ (The lower $(m,n)$-order Padé approximations are listed in Table S1). Intriguingly, even though $H_{RM}$ contains positive NN coupling terms, it is found analytically that negative NNN couplings term can already emerge in the (1,1)-order Padé approximated Hamiltonian

$$H_{eff}^{(1,1)} \approx H_{RM} - \frac{1}{2kn_0}H_{RM}^2, \quad (4)$$

in which the term $-\frac{1}{2kn_0}H_{RM}^2$ contributes a dominant negative NNN coupling, so that, $\tilde{\kappa}_{NNN} \sim -\kappa_0^2/(2kn_0)$. We also obtain the NNN ($\tilde{\kappa}_{NNN}$) term directly from the exact Helmholtz equation. The logical and pedagogical steps with $\tilde{\kappa}_{NNN}$ are described in Supplementary Materials. Therefore, the effective negative NNN couplings are qualitatively related to the intrinsic breakdown of the paraxial approximation in the microwave regime, as shown in Fig. S7 for details. Besides, when increasing the wavevector $k$ (decreasing the NN coupling $\kappa_{NN}$) to some value, for instance, entering the optical infrared regime (enlarging the waveguide spacing), the effective negative NNN couplings can be neglected, and the problem is reduced to the paraxial approximation. To further confirm the above explanation, as shown in Figs. S8 and S9, we present the simulation results of the optical analog of the pump process with a much higher frequency at near-infrared regime ($\lambda = 1.55\,\mu\text{m}$) and different waveguide spacing of the propagating electromagnetic waves. In this setup, the optical waveguide configurations are also deployed based on the Rice-Mele model, and the paraxial approximation is more strictly satisfied and the negative NNN coupling is negligible, so that the pump behaviors of both two channels appear to be symmetric as expected. Additionally, to exclude the nonadiabatic driving as the reason for asymmetry, we extend the propagation distances from 42 cm to 60 cm and 80 cm, as shown in Fig. S10, which can further confirm the nonparaxial mechanism of asymmetric pumping.

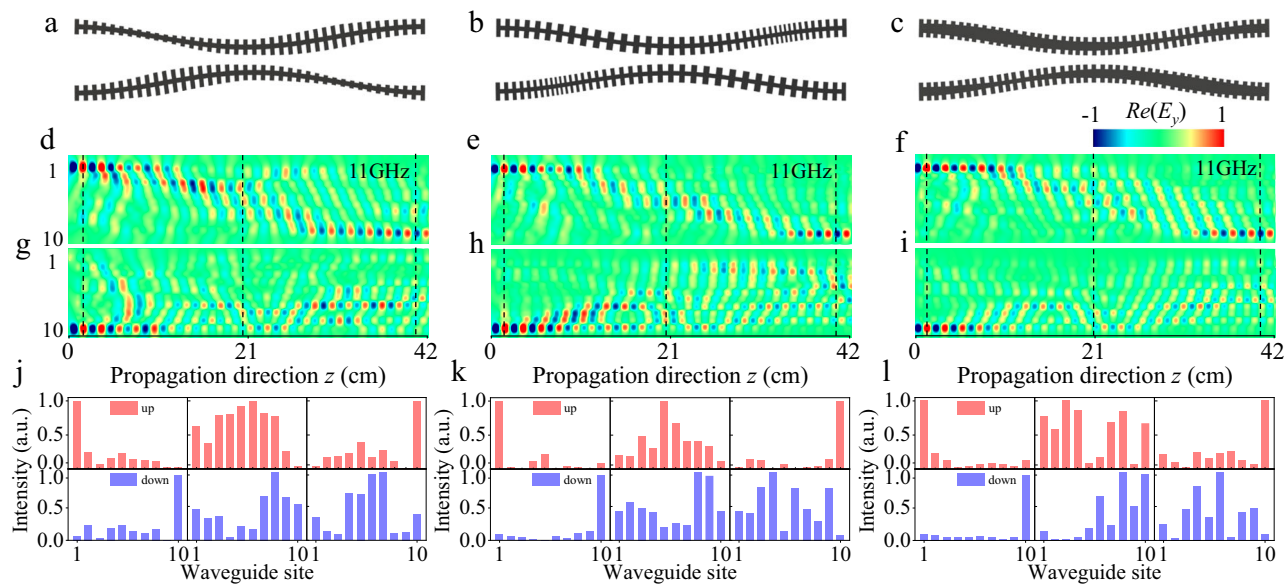

**Fig. 3 Experimental observation of asymmetric topological pumping.** Three types of photonic-lattice configurations are considered, including engineering of height, period, and girder width profiles of the "H-shaped" structures in ultra-thin corrugated metallic coupled waveguide arrays. Schematic of altering **a** height, **b** period, and **c** girder width. **d–f** Near-field evolution of the measured electric field $E_y$ at the excited frequency 11 GHz from the 1st site shows the openness of the pump channel. **g–i** Near-field evolution of the measured $E_y$ from the 10th site show the disability of the pumping. Correspondingly, the two upper and lower panels in **j**, **k**, and **l** show the instantaneous field distributions at the propagation distances $z = 0, 21, 42$ cm, as shown by the black dotted lines in **d–i**, respectively, which are regarded as the chosen phase $\phi$ in one driven period equals to 0, $\pi$, and $2\pi$, respectively.

We reveal that the asymmetric pump dynamics is ubiquitous in various waveguide designs and it should also be expected in other photonic waveguide systems where the paraxial approximation is violated. Figure 3a, b, c demonstrate three types of modulations of height, period and girder width of the "H-shaped" unit microstructure, to achieve the periodic on-site potentials in the photonic Rice-Mele model (Figs. S3 and S6 show the structural designs and experimental results measured at 17 GHz). For all the cases, the microwave field injected from the left boundary waveguide can almost be transferred to the right boundary, see Fig. 3d, e, f. However, the opposite right-to-left pump channels are blocked, in which the input microwave signal is scattered into a bulk, see Fig. 3g, h, i. For a better illustration of the difference between two pump channels of the three array designs, we plot the instantaneous intensity distributions in the arrays at the beginning ($\phi = 0$), middle ($\phi = \pi$), and end ($\phi = 2\pi$) of the pump cycle, as presented in Fig. 3j, k, l.

In summary, we have predicted the asymmetric topological pumping of light extending substantially the conventional framework of topological photonics beyond the paraxial approximation. We have verified this concept experimentally for microwave waveguide arrays, and have demonstrated that the effect of nonparaxiality can block or open the topological pump channels by altering the adiabatic passage. Our results suggest that the existing quantum-optical analogy widely used to characterize topological phases in photonics should be extended to the nonparaxial regime described by the exact Helmholtz equation. This would bring unexplored richness to the well-studied quantum simulations of topological phases in microwave photonic lattices.

### Data availability
The data supporting the findings of this study are available from the open Digital Repository at https://doi.org/10.17605/OSF.IO/58FGY.

### Code availability
All numerical codes are available from the corresponding authors on reasonable request.

### References
1. Rechtsman, M. C. et al. Photonic Floquet topological insulators. *Nature* **496**, 196 https://www.nature.com/articles/nature12066 (2013).
2. Lu, L., Joannopoulos, J. D. & Soljačić, M. Topological photonics. *Nat. Photonics* **8**, 821 https://doi.org/10.1038/nphoton.2014.248 (2014).
3. Ozawa, T. et al. Topological photonics. *Rev. Mod. Phys.* **91**, 015006 https://doi.org/10.1103/RevModPhys.91.015006 (2019).
4. Smirnova, D., Leykam, D., Chong, Y. & Kivshar, Y. Nonlinear topological photonics. *Appl. Phys. Rev.* **7**, 021306 https://doi.org/10.1063/1.5142397 (2020).
5. Segev, M. & Bandres, M. A. Topological photonics: where do we go from here? *Nanophotonics* **10**, 425 https://doi.org/10.1515/nanoph-2020-0441 (2021).
6. Hasan, M. Z. & Kane, C. L. Colloquium: topological insulators. *Rev. Mod. Phys.* **82**, 3045 https://doi.org/10.1103/RevModPhys.82.3045 (2010).
7. Xiao, D., Chang, M. C. & Niu, Q. Berry phase effects on electronic properties. *Rev. Mod. Phys.* **82**, 1959 https://doi.org/10.1103/RevModPhys.82.1959 (2010).
8. Cheng, Q., Pan, Y., Wang, Q., Li, T. & Zhu, S. Topologically protected interface mode in plasmonic waveguide arrays. *Laser Photonics Rev.* **9**, 392 https://doi.org/10.1002/lpor.201400462 (2015).
9. Blanco-Redondo, A. et al. Topological optical waveguiding in silicon and the transition between topological and trivial defect states. *Phys. Rev. Lett.* **116**, 163901 https://doi.org/10.1103/PhysRevLett.116.163901 (2016).

10. Weimann, S. et al. Topologically protected bound states in photonic parity-time-symmetric crystals. *Nat. Mater.* **16**, 433 https://doi.org/10.1038/nmat4811 (2017).
11. Wang, Z., Chong, Y. D., Joannopoulos, J. D. & Soljačić, M. Reflection-free one-way edge modes in a gyromagnetic photonic crystal *Phys. Rev. Lett.* **100**, 013905 https://doi.org/10.1103/PhysRevLett.100.013905 (2008).
12. Wang, Z., Chong, Y., Joannopoulos, J. D. & Soljačić, M. Observation of unidirectional backscattering-immune topological electromagnetic states *Nature* **461**, 772–775 https://doi.org/10.1038/nature08293 (2009).
13. Hafezi, M., Demler, E. A., Lukin, M. D. & Taylor, J. M. Robust optical delay lines with topological protection *Nat. Phys.* **7**, 907–912 https://doi.org/10.1038/nphys2063 (2011).
14. Hafezi, M., Demler, E. A., Lukin, M. D. & Taylor, J. M. Imaging topological edge states in silicon photonics *Nat. Photon.* **7**, 425 https://doi.org/10.1038/nphoton.2013.274 (2013).
15. Thouless, D. J., Kohmoto, M., Nightingale, M. P. & den Nijs, M. Quantized Hall conductance in a two-dimensional periodic potential. *Phys. Rev. Lett.* **49**, 405 https://doi.org/10.1103/PhysRevLett.49.405 (1982).
16. Kraus, Y. E., Lahini, Y., Ringel, Z., Verbin, M. & Zilberberg, O. Topological states and adiabatic pumping in quasicrystals. *Phys. Rev. Lett.* **109**, 629 https://doi.org/10.1103/PhysRevLett.109.106402 (2012).
17. Verbin, M., Zilberberg, O., Kraus, Y. E., Lahini, Y. & Silberberg, Y. Observation of topological phase transitions in photonic quasicrystals. *Phys. Rev. Lett.* **110**, 076403 https://doi.org/10.1103/PhysRevLett.110.076403 (2012).
18. Kraus, Y. E., Ringel, Z. & Zilberberg, O. Four-dimensional quantum Hall effect in a two-dimensional quasicrystal. *Phys. Rev. Lett.* **111**, 226401 https://doi.org/10.1103/PhysRevLett.111.226401 (2013).
19. Verbin, M., Zilberberg, O., Lahini, Y., Kraus, Y. E. & Silberberg, Y. Topological pumping over a photonic fibonacci quasicrystal. *Phys. Rev. B* **91**, 064201 https://doi.org/10.1103/PhysRevB.91.064201 (2014).
20. Zilberberg, O. et al. Photonic topological boundary pumping as a probe of 4d quantum Hall physics. *Nature* **553**, 59 https://doi.org/10.1038/nature25011 (2018).
21. Lohse, M., Schweizer, C., Price, H. M., Zilberberg, O. & Bloch, I. Exploring 4d quantum Hall physics with a 2d topological charge pump. *Nature* **553**, 55 https://doi.org/10.1038/nature25000 (2018).
22. St-Jean, P. et al. Lasing in topological edge states of a one-dimensional lattice. *Nat. Photonics* **11**, 651 https://doi.org/10.1038/s41566-017-0006-2 (2017).
23. Zhao, H. et al. Topological hybrid silicon microlasers. *Nat. Commun.* **9**, 981 https://doi.org/10.1038/s41467-018-03434-2 (2018).
24. Harari, G. et al. Topological insulator laser: theory. *Science* **359**, eaar4003 https://doi.org/10.1126/science.aar4003 (2018).
25. Bandres, M. A. et al. Topological insulator laser: experiments. *Science* **359**, eaar4005 https://doi.org/10.1126/science.aar4005 (2018).
26. Zeng, Y. et al. Electrically pumped topological laser with valley edge modes. *Nature* **578**, 246 https://doi.org/10.1038/s41586-020-1981-x (2020).
27. Kim, H.-R. et al. Multipolar lasing modes from topological corner states. *Nat. Commun.* **11**, 5758 https://doi.org/10.1038/s41467-020-19609-9 (2020).
28. Song, W. et al. Breakup and recovery of topological zero modes in finite non-Hermitian optical lattices. *Phys. Rev. Lett.* **123**, 165701 https://doi.org/10.1103/PhysRevLett.123.165701 (2019).
29. Zhou, X., Wang, Y., Leykam, D. & Chong, Y. D. Optical isolation with nonlinear topological photonics. *N. J. Phys.* **19**, 095002 https://doi.org/10.1088/1367-2630/aa7cb5 (2017).
30. Chen, W., Leykam, D., Chong, Y. & Yang, L. Nonreciprocity in synthetic photonic materials with nonlinearity. *MRS Bull.* **43**, 443 https://doi.org/10.1557/mrs.2018.124 (2018).
31. Kruk, S. et al. Nonlinear light generation in topological nanostructures. *Nat. Nanotechnol.* **14**, 126 https://doi.org/10.1038/s41565-018-0324-7 (2019).
32. Wang, Y., Lang, L.-J., Lee, C. H., Zhang, B. & Chong, Y. Topologically enhanced harmonic generation in a nonlinear transmission line metamaterial. *Nat. Commun.* **10**, 1102 https://doi.org/10.1038/s41467-019-08966-9 (2019).
33. Cerjan, A., Wang, M., Huang, S., Chen, K. P. & Rechtsman, M. C. Thouless pumping in disordered photonic systems. *Light Sci. Appl.* **9**, 178 https://doi.org/10.1038/s41377-020-00408-2 (2020).
34. Fedorova, Z., Qiu, H., Linden, S. & Kroha, J. Observation of topological transport quantization by dissipation in fast Thouless pumps. *Nat. Commun.* **11**, 3758 https://doi.org/10.1038/s41467-020-17510-z (2020).
35. Marra, P. & Nitta, M. Topologically quantized current in quasiperiodic Thouless pumps. *Phys. Rev. Res.* **2**, 042035(R) https://doi.org/10.1103/PhysRevResearch.2.042035 (2020).
36. Blanco-Redondo, A., Bell, B., Oren, D., Eggleton, B. J. & Segev, M. Topological protection of biphoton states. *Science* **362**, 568 https://doi.org/10.1126/science.aau4296 (2018).
37. Tambasco, J.-L. et al. Quantum interference of topological states of light. *Sci. Adv.* **4**, eaat3187 https://doi.org/10.1126/sciadv.aat3187 (2018).
38. Hu, S., Ke, Y. & Lee, C. Topological quantum transport and spatial entanglement distribution via a disordered bulk channel. *Phys. Rev. A* **101**, 052323 https://doi.org/10.1103/PhysRevA.101.052323 (2020).
39. Ke, Y., Qin, X., Kivshar, Y. & Lee, C. Multiparticle Wannier states and Thouless pumping of interacting bosons. *Phys. Rev. A* **95**, 063630 https://doi.org/10.1103/PhysRevA.95.063630 (2017).
40. Wang, Y. et al. Topological protection of two-photon quantum correlation on a photonic chip. *Optica* **6**, 955 https://doi.org/10.1364/OPTICA.6.000955 (2019).
41. Wang, Y. et al. Quantum topological boundary states in quasi-crystals. *Adv. Mater.* **31**, 1905624 https://doi.org/10.1002/adma.201905624 (2019).
42. Thouless, D. J. Quantization of particle transport. *Phys. Rev. B* **27**, 6083 https://doi.org/10.1103/PhysRevB.27.6083 (1983).
43. Switkes, M., Marcus, C., Campman, K. & Gossard, A. An adiabatic quantum electron pump. *Science* **283**, 1905 https://doi.org/10.1126/science.283.5409.1905 (1999).
44. Ho, D. Y. H. & Gong, J. Quantized adiabatic transport in momentum space. *Phys. Rev. Lett.* **109**, 010601 https://doi.org/10.1103/PhysRevLett.109.010601 (2012).
45. Lohse, M., Schweizer, C., Zilberberg, O., Aidelsburger, M. & Bloch, I. A Thouless quantum pump with ultracold bosonic atoms in an optical superlattice. *Nat. Phys.* **12**, 350 https://doi.org/10.1038/nphys3584 (2016).
46. Nakajima, S. et al. Topological Thouless pumping of ultracold fermions. *Nat. Phys.* **12**, 296 https://doi.org/10.1038/nphys3622 (2016).
47. Ma, W. et al. Experimental observation of a generalized Thouless pump with a single spin. *Phys. Rev. Lett.* **120**, 120501 https://doi.org/10.1103/PhysRevLett.120.120501 (2018).
48. Ke, Y. et al. Topological phase transitions and Thouless pumping of light in photonic waveguide arrays. *Laser Photonics Rev.* **10**, 995 https://doi.org/10.1002/lpor.201600119 (2016).
49. Wauters, M. M., Russomanno, A., Citro, R., Santoro, G. E. & Privitera, L. Localization, topology, and quantized transport in disordered Floquet systems. *Phys. Rev. Lett.* **123**, 266601 https://doi.org/10.1103/PhysRevLett.123.266601 (2019).
50. Petrides, I. & Zilberberg, O. Higher-order topological insulators, topological pumps and the quantum Hall effect in high dimensions. *Phys. Rev. Res.* **2**, 022049 https://doi.org/10.1103/PhysRevResearch.2.022049 (2020).
51. Benalcazar, W. A. et al. Higher-order topological pumping, arXiv:2006.13242 (2020)
52. Kolodrubetz, M. H., Nathan, F., Gazit, S., Morimoto, T. & Moore, J. E. Topological Floquet-Thouless energy pump. *Phys. Rev. Lett.* **120**, 150601 https://doi.org/10.1103/PhysRevLett.120.150601 (2018).
53. Hu, S., Ke, Y., Deng, Y. & Lee, C. Dispersion-suppressed topological Thouless pumping. *Phys. Rev. B* **100**, 064302 https://doi.org/10.1103/PhysRevB.100.064302 (2019).
54. Longhi, S. Topological pumping of edge states via adiabatic passage. *Phys. Rev. B* **99**, 155150 https://doi.org/10.1103/PhysRevB.99.155150 (2019).
55. Ke, Y. et al. Topological pumping assisted by Bloch oscillations. *Phys. Rev. Res.* **2**, 033143 https://doi.org/10.1103/PhysRevResearch.2.033143 (2020).
56. Penjweini, R., Weber, M., Sondermann, M., Boyd, R. W. & Leuchs, G. Nonlinear optics with full three-dimensional illumination. *Optica* **6**, 878 https://doi.org/10.1364/OPTICA.6.000878 (2019).
57. Rice, M. J. & Mele, E. J. Elementary excitations of a linearly conjugated diatomic polymer. *Phys. Rev. Lett.* **49**, 1455 https://doi.org/10.1103/PhysRevLett.49.1455 (1982).
58. Ronald Hadley, G. Wide-angle beam propagation using Padé approximant operators. *Opt. Lett.* **17**, 1426 https://doi.org/10.1364/OL.17.001426 (1992).
59. Shen, S. Q., *Topological Insulators: Dirac Equation in Condensed Matters* (Springer, 2017).
60. Cheng, Q. et al. Observation of anomalous $\pi$ modes in photonic Floquet engineering. *Phys. Rev. Lett.* **122**, 173901 https://doi.org/10.1103/PhysRevLett.122.173901 (2019).
61. Ma, H. F., Shen, X., Cheng, Q., Jiang, W. X. & Cui, T. J. Broadband and high-efficiency conversion from guided waves to spoof surface plasmon polaritons. *Laser Photonics Rev.* **8**, 146 https://doi.org/10.1002/lpor.201300118 (2014).
62. Cheng, Q. et al. Flexibly designed spoof surface plasmon waveguide array for topological zero-mode realization. *Opt. Express* **26**, 31636 https://doi.org/10.1364/OE.26.031636 (2018).
63. Chen, Z.-G., Tang, W., Zhang, R.-Y., Chen, Z. & Ma, G. Landau-Zener transition in the dynamic transfer of acoustic topological states. *Phys. Rev. Lett.* **126**, 054301 https://doi.org/10.1103/PhysRevLett.126.054301 (2021).

## Acknowledgements

Q. Cheng is supported by the National Natural Science Foundation of China (grant 11874266, 12174260), by the Shanghai Rising-Star Program (grant 21QA1406400) and by the Shanghai Science and Technology Development Fund (grant 21ZR1443500). H. Wang is supported by the National Natural Science Foundation of China (grant 12104217). Y. Ke is supported by the National Natural Science Foundation of China (grant 11904419). C. Lee is supported by the National Natural Science Foundation of China (grant 12025509 and 11874434), the Key-Area Research and Development Program of GuangDong Province (grant 2019B030330001), and the Science and Technology Program of Guangzhou (grant 201904020024). Y. Kivshar is supported by the Australian Research Council (grant DP200101168).

## Author contributions

Q.C., H.W., and Y. Ke contributed equally to this work. Q.C., H.W., and Y.P. conceived the idea. Q.C., Y.Y., and T.C. designed the microwave waveguide arrays and performed the numerical simulations. Q.C., Y.Y., and T.C. designed the experiments and fabricated the sample. Y.Y. and T.C. conducted the measurements. Y. Ke and H.W. carried out the theoretical calculations. Y.K., C.L., and Y.P. supervised the project. All authors contributed extensively to the interpretation of the results and the writing of the paper.

## Competing interests

The authors declare no competing interests.

## Additional information

**Supplementary information** The online version contains supplementary material available at https://doi.org/10.1038/s41467-021-27773-9.

**Correspondence** and requests for materials should be addressed to Yuri S. Kivshar, Chaohong Lee or Yiming Pan.

**Peer review information** *Nature Communication* thanks Baile Zhang and the other, anonymous, reviewer(s) for their contribution to the peer review of this work.

**Reprints and permission information** is available at http://www.nature.com/reprints

**Publisher's note** Springer Nature remains neutral with regard to jurisdictional claims in published maps and institutional affiliations.

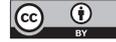